\documentclass[twocolumn,showpacs,superscriptaddress,groupedaddress]{revtex4}
\usepackage[colorlinks=true, linkcolor=red, citecolor=blue, urlcolor=blue]{hyperref}
\usepackage[utf8]{inputenc}
\usepackage{graphicx}
\usepackage{algorithm}
\usepackage{algpseudocode}
\usepackage{amssymb}
\usepackage{amsmath}
\usepackage{dsfont}
\usepackage{tikz-cd}
\usepackage{enumitem}
\usepackage[percent]{overpic}
\setenumerate[1]{label=(\arabic*)}

\newcommand{\ket}[1]{\left|{#1}\right\rangle}
\newcommand{\bra}[1]{\left\langle{#1}\right|}
\newcommand{\xddots}{%
  \raise 7pt \hbox {.}
  \mkern 3mu
  \raise 3pt \hbox {.}
  \mkern 3mu
  \raise -1pt \hbox {.}
}

\begin{document}
\title{Epistemic Horizons and the Foundations of Quantum Mechanics}

\author{Jochen Szangolies}
\email{Jochen.Szangolies@gmx.de}

%

\begin{abstract}

In-principle restrictions on the amount of information that can be gathered about a system have been proposed as a foundational principle in several recent reconstructions of the formalism of quantum mechanics. However, it seems unclear precisely why one should be thus restricted. We investigate the notion of paradoxical self-reference as a possible origin of such epistemic horizons by means of a fixed-point theorem in Cartesian closed categories due to F. W. Lawvere that illuminates and unifies the different perspectives on self-reference.

\end{abstract}
\pacs{03.65.-w, 03.65.Ta}

\maketitle

\section{Introduction}

`Why the Quantum?' is one of John Wheeler's `Really Big Questions' \cite{bar2004}, and possibly the one that has received the most attention in print. It invites looking at quantum mechanics not from an interpretive, but rather, from a reconstructive point of view: what, exactly, is the reason that nature is described by quantum mechanical laws? 

Several authors have interpreted this question as asking for a \textit{foundational principle}, that is, a short (and ideally, intuitive and compelling) statement whose truth implies the necessity of the quantum formalism. Special relativity has such a principle in the invariant speed of light; general relativity has it in the equivalence principle. But for quantum mechanics, such a foundation seems absent.

In recent years, however, a popular candidate that has been advanced is an \textit{epistemic horizon} of some sort. By an epistemic horizon, I mean a fundamental limit to the knowledge available about a system to any conceivable observer. Grinbaum \cite{Gr2003} identifies two basic principles as common across several different approaches towards a first-principles reconstruction of quantum mechanics, due to Rovelli \cite{Rov1996}, Brukner and Zeilinger \cite{BZ2003}, and Fuchs \cite{Fu2002}. Similar principles are at work in the reconstructions due to Masanes et al. \cite{MM2013} and H\"ohner and Wever \cite{HW2015}. Finally, an early example is provided by von Weizs\"acker's reconstruction of quantum theory in terms of what he calls `ur-alternatives' (where `ur' means, roughly, primitive or primordial) \cite{vW1985}.

These two principles can be summarised as:
\begin{enumerate}
 \item \label{fin}\textit{Finiteness assumption}: There is a finite maximum of information that can be obtained about any given system. 
 \item \label{add}\textit{Assumption of additional information}: It is always possible to acquire new information about any system.
\end{enumerate}

At first blush, the two assumptions above appear to be contradictory: if the state of the system is maximally known, no further information can in principle be obtained anymore. That this is not so is demonstrated by a related, but distinct approach due to Spekkens \cite{Sp2007}, who proposes a `toy theory', i.e. a theory capable of reproducing some, but not all, of the phenomena of quantum mechanics, based on the \textit{knowledge balance principle}:
\begin{quotation}
 ``If one has maximal knowledge, then for
every system, at every time, the amount of
knowledge one possesses about the ontic state
of the system at that time must equal the
amount of knowledge one lacks.''
\end{quotation}
The toy theory serves as a useful means of illustrating the precise role epistemic horizons play regarding those phenomena that are usually considered to be distinctly `quantum', including entanglement, complementary measurements, teleportation, interference, and the impossibility of cloning an arbitrary state. It is therefore instructive to take a look at the way these phenomena are produced in the toy theory.

First, an elementary system in the toy theory, called a `toy bit', is specified by two bits of information---however, due to the knowledge balance principle, only one of them can be known at any given time. The toy bit thus has four possible states, denoted $t_1,t_2,t_3$ and $t_4$, and maximal achievable knowledge of its state suffices to `localize' it within any two-element subset of the state space. That is, there are six toy bit states of maximum knowledge, $\{t_1,t_2\}$, $\{t_2,t_3\}$, $\{t_1,t_3\}$, $\{t_2,t_4\}$, $\{t_1,t_4\}$, $\{t_3,t_4\}$, and one single state of less-than-maximal knowledge, here equivalent to the state of complete ignorance $\{t_1,t_2,t_3,t_4\}$. In order to gauge the importance of the epistemic horizon in this theory, we will have a brief look at the phenomenon of complementarity. 

Measurements in the toy theory are such that they distinguish between two disjunct subsets of the state space. Thus, they can be considered to be the three $\{0,1\}$-valued functions specified in Table~\ref{measTT}.

\begin{center}
 \begin{table}[h]
  \caption{Measurements in Spekkens' toy theory.} 
  \begin{tabular}{c | c c c c}\label{measTT}
             & $t_1$ & $t_2$ & $t_3$ & $t_4$ \\ \hline
  $m_z$  & 1 & 1 & 0 & 0 \\
  $m_x$  & 1 & 0 & 1 & 0 \\
  $m_y$  & 1 & 0 & 0 & 1 \\
  \end{tabular}
 \end{table}
\end{center}

As the indices suggest, these can be considered to be the analogues of the three mutually orthogonal measurement directions on a qubit. This means that the state space of a toybit is an octahedron, with the six maximal knowledge states on the vertices, and the single maximally mixed state at the center, providing a discrete analogue to the Bloch ball.

Let us now assume we perform successive measurements of $m_z$ and $m_x$. Let the outcome of the first measurement be $1$. Then, it is clear that the outcome of the second measurement is not determined: perfect knowledge of the outcome of the $m_z$ measurement entails ignorance of the $m_x$ outcome. 

However, let the outcome of the second $m_x$ measurement now likewise be $1$. Since the only state compatible with both outcomes is $t_1$, we would be able to deduce, in conflict with the knowledge balance principle, that the state of the toy bit now is $t_1$. To prevent this, the state must be changed after the second measurement, in such a way as to be not necessarily compatible with $m_z(t)=1$ anymore, but still with $m_x(t)=1$ (as otherwise, one could hardly speak of measurement). A re-measurement of $m_z$ will then again equiprobably yield either of its outcomes: $m_z$ and $m_x$ are complementary.

The toy theory hence possesses the feature of novel information (albeit a finite amount) always being available, despite having maximal information about the system---thus showing that the two principles are not contradictory, but are instead responsible for the emergence of many non-trivial analogues of quantum effects.

This situation is in fact not unlike that of an actual horizon: we only ever see a small section of the Earth's surface, due to its curvature. Attempting to move closer to the boundary of our knowledge---the horizon---results in novel features of the landscape being revealed, at the cost of losing sight of what we left behind. Hence, there is a maximum of information about the Earth's surface available to us, yet we can gather novel information at will. 

As we have seen, quantum-like features flow relatively immediately from these epistemically restrictive approaches. An intuitive reason for this is that a limit on the information obtainable from a system precludes it being exactly localizable in phase space (or some more general state space): there will be a minimum phase space volume, within which the system cannot be further localized. As Rovelli notes \cite{Rov1996}, this limitation already introduces the constant $\hbar$ into the description: this minimum volume will have dimensions of $(\mathrm{kgm^2s^{-1}})^{3n}$ for a system composed of $n$ particles, implying the existence of a constant of maximum localizability with dimensions $\mathrm{kgm^2s^{-1}}$, i.e. $\hbar$. This also provides an intuitive reason for the uncertainty relations.

The existence of such a minimum phase-space volume deforms the (classical) algebra of functions on phase space into the non-commutative Moyal algebra, which is in one-to-one correspondence with the algebra of operators on Hilbert space in the traditional formulation of quantum mechanics. This yields the basis for the so-called \textit{deformation quantization} (for a review, see e.g. \cite{CZ2012}). 

A further observation made by Grinbaum (\cite{Gr2005}) is the fact that imposing a limit on the information available about a system imposes a logical calculus similar to the quantum logic of Birkhoff and Neumann (given by an orthomodular lattice) \cite{Bir1936}. 

This brief review demonstrates that epistemic horizons are a promising candidate for a foundational principle of quantum mechanics, explaining at least some of its properties in a natural way. Our aim in this article is then to investigate a possible origin for such restrictions. Our first cue towards this end is the appearance of randomness in quantum mechanics. There is a close connection between the principles \ref{fin} and \ref{add} and the notion of randomness: if the state of a system is maximally known according to \ref{fin}, any additional information \ref{add} assures us can always be acquired cannot be reducible to this prior knowledge, and hence, must be random.

However, in mathematics, randomness is closely tied to undecidability---the existence of propositions that cannot be decided within a given axiomatic framework \cite{Ch1990}. 

Thus, in Sec.~\ref{selfref} we will consider the connection between undecidability and randomness in more detail, focusing on some approaches towards applying self-referential arguments to physical systems. Afterwards, in Sec.~\ref{main}, we will consider physical self-reference in a novel setting, which we will use to establish the first of our main results, proving that there exists no specification of a system's state such that the outcomes of all further experiments are deducible from it. This gives a concrete foundation for the second principle above, namely, that new information about a system can always be acquired. In the following Sec.~\ref{quant}, using an argument from algorithmic information theory, our second main result is that a given system can only be localized to a finite degree of accuracy within its state space, likewise justifying the principle of maximum information, and establishing the necessity of epistemic horizons. In Section~\ref{conc}, we conclude and give an outlook towards future research extending the present work. 

\section{Self-Reference in Physics} \label{selfref}

On most standard accounts, quantum mechanical measurement produces outcomes that may be random---and moreover, whose randomness is both \textit{irreducible} and \textit{genuine}. That is, the randomness is neither due to our ignorance about the system, nor is it a product of underlying deterministic (perhaps chaotic) processes. Indeed, Yurtsever has argued that quantum mechanical randomness must be genuine---otherwise, for a sufficiently long measurement sequence, any correlations present in pseudorandom data would allow nonlocal signalling between spacelike separated agents \cite{Yu1998}. Likewise, Bendersky et al.~\cite{ben2017} show that any deterministic theory reproducing the predictions of quantum mechanics must be noncomputable, or else, lead to exploitable signaling. 

A similar conclusion is reached by Calude and Svozil \cite{CS2008}, who demonstrate that the value indefiniteness (the nonexistence of simultaneous definite values for all observables) suggested by the Kochen-Specker theorem \cite{KS1967} and related results implies the incomputability of quantum randomness. 

This has an immediate, but perhaps underappreciated, corollary: namely, no algorithmic process can account for all measurements outcomes---some measurements, hence, must be algorithmically undecidable. In the words of Feynman \cite{Fe1982}:
\begin{quotation}
 ``It is impossible to represent the results of quantum mechanics with a classical universal device."
\end{quotation}

Of special interest here is a result due to Edis (\cite{Edi1998}), who proves that every incomputable function can be represented by a finite algorithm augmented by an infinite lookup table of random bits. Hence, every such function can be decomposed into an algorithmic and a random part. This suggests that systems behaving according to a noncomputable law show a behavior that is algorithmic, interspersed with random events---which strongly recalls the two-tiered dynamics familiar from quantum theory.

\begin{algorithm}
\caption{Algorithm augmented with infinite random data.}\label{QMA}
\begin{algorithmic}[1]
\Procedure{u}{$n$}
\If{$n\,\mathrm{mod}\,3 \neq 0$}

\Return{$n^2$}

\Else

\Return{$\rho(n)$}

\EndIf
\EndProcedure
\\
\State $\rho \gets 1101110101011010\ldots$
\end{algorithmic}
\end{algorithm}

Thus, as in the example, a noncomputable law might yield random data under certain conditions, and otherwise work according to a deterministic law (the above algorithm computes the square of all inputs not divisible by three, and returns a random bit otherwise). Observation of the output of this `algorithm' then allows one to deduce the deterministic part, while leaving the rest to appear random.

This remark deserves some further clarification. Due to the fact that we observe any physical system only for a finite time, it is trivially the case that there exists a Turing machine, or algorithm, replicating its behavior. However, in general, the `law' thus found will not be significantly less complex than the behavior itself, in the same way that an algorithm reproducing a finite random number will be of a length approximately equal to that number (if the number is $x$, the algorithm might simply be `print $x$'). Furthermore, it will not allow any prediction of the future behavior of the system.

Decomposing the system's behavior into an algorithmic and a random part might then afford at least partial compression of the system's behavior, and allow prediction in at least a subset of cases. Consequently, an observer might well arrive at such a `composite' description for a system behaving according to a noncomputable law.

There exists another way to highlight the deep mathematical connection between randomness and undecidability. Consider the following construction, known as \textit{Chaitin's constant} \cite{Ch1975}:
\begin{equation}
 \Omega_U=\sum_{x:U(x)\,\mathrm{halts}}2^{-|x|}. 
\end{equation}

Here, $U$ is a given (universal) Turing machine, with the special property that all valid programs for $U$ self-terminate, that is, if $x$ is a valid program, there is no $x^\prime$ such that it extends $x$ and is also a valid program. Programs $x$ can be thought of as being represented by bit strings; $|x|$ then denotes their length in bits. Thus, $\Omega_U$ yields the probability that $U$ halts, given a random program, and is hence also referred to as \textit{halting probability}. 

Knowledge of the first $n$ bits of the halting probability of $U$ entails being able to solve the halting problem for $U$ for all programs of length $n$ \cite{Ch1975}: one simply runs all programs of length up to $n$ bits in parallel, and adds the lengths of those that have halted to the estimate for $\Omega_U$; once this estimate matches the first $n$ bits, the programs that have not yet halted will never halt. Due to the recursive insolubility of the halting problem \cite{Tu1936}, $\Omega_U$ must be uncomputable, and thus, the values of each of its bits represent undecidable propositions.

Furthermore, $\Omega_U$ must be a random number; in fact, it can be shown that any recursively enumerable number---a number that is the limit of a computable, increasing series of rational numbers---that is random is a halting probability \cite{Ca2001}. Hence, instances of randomness in mathematics are closely tied to undecidable propositions, and wherever we find randomness, we should look for undecidability at its core. 

For a more in-depth look at the connections between randomness and undecidability, as well as their implications for physics, see the books by Svozil \cite{Sv1993,svo2018}.

At the outset, it may not be immediately clear how the notion of undecidability applies to physical systems. The fact that for any given axiom system (of sufficient strength), there exist propositions that the system cannot decide was first pointed out by G\"odel \cite{Go1931}, in a paper that famously overturned Hilbert's hope of finding a single axiom system encompassing all of mathematics. Shortly thereafter, Turing \cite{Tu1936} proposed what is to this day the canonical example of an undecidable problem, namely, the question of whether a given algorithm ever halts and produces an output. 

Both of these results apply strictly in the formal setting of mathematical logic. However, they essentially rely on the notion of self-reference: Gödel's theorem applies to all formal systems capable of axiomatizing (a certain fraction of) number theory, since such theories are capable of formalizing statements about themselves, while Turing's proof relies on constructing a program capable of solving its own halting problem. Both essentially frame limitations by showing the paradoxical consequences of certain self-referential constructions: a theorem that is provable if and only if it is not, or a program that halts if and only if it fails to. In this way, paradoxical self-reference produces epistemic horizons in the form of undecidable propositions. However, self-reference is not limited to the setting of mathematical logic; thus, analogues to these phenomena may be found in other domains.

Indeed, in his proof, Turing introduced one of the first general formalizations of what it means to compute a function, via the notion of what he called \textit{a-machines} (now usually called \textit{Turing machines} in his honor). This ties the formal argumentation to a concrete system that can, in principle, be instantiated physically---and indeed, modern-day computers, if outfitted with an unbounded supply of data storage, are equivalent to Turing machines in their computational power. Thus, the question of whether a given algorithm ever halts can be reformulated as a question of whether a computer ever reaches a certain pre-defined halting state, hence translating the formal question to a question about the dynamical evolution of a concrete physical system (which thus must be undecidable). 

There exists a broad analogy between the time-evolution of physical systems, the act of deriving theorems in formal systems, and computation: in each case, we start of with some initial information---the formal system's axioms, the initial state of the physical system, or the program---and apply a transformation in order to derive new information: a new theorem, the state of the system at a point in the future, or the computation's output. 

Indeed, we benefit from this analogy every day: thanks to it, we may view certain physical system as performing computations. Likewise, for any formal system, there exists a Turing machine that enumerates its theorems; more generally, the equivalence between proofs and programs is established via the Curry-Howard correspondence \cite{Ho1980}. 

A distinction between these three domains might seem to be the fact that theorem-deriving and computation may be information-lossy: the output of a computation does not necessarily allow the unique reconstruction of its initial data, and a theorem may not imply the full set of axioms of a formal system. Conversely, a physical system evolves in an information-preserving way, such that the dynamics may be reversed to recover the initial state.

However, this distinction is ultimately superficial: we might make computation information-preserving by exclusively considering reversible computation (which has the same power as ordinary computation \cite{Be1973}), or we might restrict our attention to only a part of a physical system, integrating out certain degrees of freedom, to implement information loss. On the side of formal systems, the derivation of a theorem---being at each step reducible to pure syntactical manipulations---does not introduce additional information, but uncovers implications of the axioms. This is the origin of \textit{Chaitin's principle} \cite{Ch1982}, stating roughly that it is impossible to derive a theorem from a set of axioms if it is more complex than those atoms.

In fact, it is exactly this conservation of information that makes the connection between the three domains meaningful: we may consider the processes of computation, derivation and physical evolution as acting on the information contained in the initial data to `unpack' the final data.

Both the limitations on computation and logical derivation can be considered to stem from this conservation of information: Chaitin's principle implies that certain theorems cannot be derived from a set of axioms, and Edis' result shows that if we were to augment a Turing machine with a source of new information in the form of an infinite random bit string, it would be possible to `compute' uncomputable functions. Consequently, it seems natural to investigate whether there is a similar epistemic horizon in the domain of physics.

\begin{center}
 \begin{table}[h]
  \caption{Correspondence between computation, formal systems, and physics.} 
  
  \begin{tabular}{ c || c | c | c }\label{ana}
   						& Computation 		& Formal Systems 	& Physics \\ \hline \hline
   initial data: 		& program / input 	& axioms 			& initial state \\ \hline
   process: 			& computation 		& derivation 		& dynamics \\ \hline
   final data: 			& output 			& theorem 			& final state \\ \hline
   epistemic horizon: 	& uncomputability 	& incompleteness 	& ?  
              
  \end{tabular}
 \end{table}
\end{center}

\subsection{Prior Work}

With the possibility of real-word instantiations of self-reference given by the correspondence in Table~\ref{ana}, the application of undecidability results to physical systems becomes feasible. In this section, we provide a brief (and partial) overview of results obtained by exploiting this connection. 

Recently, Cubitt et al. \cite{Cu2015} have used this correspondence---between the ground state configuration of a spin system and the problem of tiling the plane with tiles whose edges are colored, such that the same colors meet, which is known to be undecidable \cite{Be1966}---to show that, in general, the question of whether this ground state possesses an energy gap is undecidable (see also \cite{Ll1993,Ll1994}). Since this energy gap is in principle a measurable quantity, the outcome of this measurement is undecidable.

Likewise, Eisert et al. \cite{Ei2012} have shown by a reduction to the undecidable matrix mortality problem---which asks whether one can multiply matrices from a given set, such as to obtain the zero matrix---that the question of whether a certain outcome in a sequence of Stern-Gerlach measurements ever occurs is undecidable.

In fact, the application of self-referential constructions to physical systems has a long history. Popper, in 1950, was one of the earliest to consider this connection \cite{Pop1950}, giving an argument establishing that no physical system can ever perfectly predict its own behaviour. In 1964, Rothstein considered a Maxwell's-demon type of setup, proposing an intrinsic limit on the accuracy of measurements that can be performed \cite{Rot1964}. 

A more daring proposal is what one might call the `Gödelian hunch': the idea that the origin of the peculiarities surrounding quantum theory lie in phenomena related, or at least similar, to that of incompleteness in formal systems. 

There is a potential for misunderstanding here, that has at times given rise to misguided criticism of the general idea: it is not proposed that quantum indeterminacy is due, in any direct sense, to Gödel's theorem. The incompleteness theorem applies to formal systems capable of axiomatizing a certain fragment of arithmetic, not to physical systems. Rather, the idea is that similar phenomena in the domain of physics, perhaps working on related principles, might be foundational to quantum mechanics in one way or another. 

This idea was given an early formulation by Wheeler himself: in unpublished notes to a discussion held with, among others, Roger Penrose and Simon Kochen, he identifies the point of origin of the `quantum principle' as the undecidable propositions of mathematical logic. (The full text of this note is reprinted in the appendix of Ref.~\cite{Fu2016}; a scanned pdf copy can be found at the URL in Ref.~\cite{whe1974}.) However, in an oft-quoted story, Wheeler reports being thrown out of G\"{o}del's office by the latter after confronting him with the suggestion of a connection between the incompleteness theorems and quantum mechanics (see e. g. Ref.~\cite{ber1991}).

Along similar lines, Dalla Chiara in 1977 \cite{DC1977} considered the importance of self-reference for the quantum measurement problem. Continuing this investigation, but extending it to the classical domain, in 1995 Breuer proved \cite{Br1995} that no observer can distinguish all the states of a system that includes them as a proper part, which he explicitly connected to the quantum measurement problem in 1999 \cite{Bre1999}. Within a similar framework, Aerts established in 2005 \cite{Ae2005} that there are properties pertaining to the observer that they cannot perfectly observe.

Yet more directly, Zwick \cite{MZ1978} has proposed an analogy between quantum measurement and paradoxical self-reference, using a discretized version of the quantum formalism that models quantum evolution as a succession of wave functions---analogous to steps of a derivation---with the result of a measurement taking the role of an undecidable proposition. A related theme was pursued by Peres and Zurek \cite{Per1982}, who argue that ``[quantum theory's] inability to completely describe the measurement process appears to be not a flaw of the theory but a \textit{logical necessity} which is analogous to G\"odel's undecidability theorem [emphasis in the original]".

The proposal by Brukner \cite{Bru2010} and Paterek et al. \cite{Pat2010} comes very close to realizing the `Gödelian hunch'. Essentially, they use the correspondence shown in Table~\ref{ana} to embed a simple set of axioms within the initial state of a quantum system, and associate a proposition with a measurement, which yields random results if and only if that proposition is independent from the axioms.

This bears a remarkable resemblance to Spekkens' knowledge balance principle: the information related to the measurement outcome cannot be derived from the axioms; yet, an outcome is produced, with the consequence that the state is changed (the `collapse' of the wave function) such that the post-measurement state contains the same amount of information as the initial state.

They indeed relate their observations to Chaitin's algorithmic information theory-based version of G\"odel's incompleteness theorem, stating informally that no theorem can be derived from a set of axioms if it contains more information than them, as quantified via an appropriate measure \cite{CJ2005}. However, they note explicitly that their axiom systems do not meet the formal requirements for the applicability of G\"odel's theorem, being unable to capture the required features of arithmetic, and thus, are completable.

Finally, Calude and Stay \cite{CS2007} establish a formal uncertainty principle between the value of a random real number and the (knowledge of the) length of the shortest program computing it that is equivalent to Chaitin's incompleteness theorem, and show that these uncertainties can be thought of as standard deviations of canonically conjugate measurable quantities for certain physically implemented (quantum-) computations. Nevertheless, the result remains partial: it is far from clear whether any instance of uncertainty in quantum mechanics can be similarly related to mathematical undecidability.

\subsection{Preliminaries}

These results point to the possibility of an intriguing connection between fundamental features of quantum mechanics and the phenomena of paradoxical self-reference. However, they remain partial. In our investigation, we will pursue the same path, and study, from a broad vantage point, the implications of self-reference for physical theories in general.

This angle of attack is justified by the fact that any universal theory---that is, any theory whose domain of validity is supposed to cover all physical systems indiscriminately, in particular, both observers and the systems they observe---must contain self-referential elements: it must describe both the properties of a given object system, as well as an observer's knowledge about these properties (which must, after all, be stored in the observer's state in some way). Hence, an investigation into the effects of such self-reference seems a natural avenue of inquiry.

First, we will introduce some notation. The mathematical representative of a physical system $\mathcal{S}$ is its \textit{state space} $\Sigma_\mathcal{S}$. In a sense to be clarified below, the state space can be considered to be a collection of all the possible ways for a physical system to be, defining its possible properties. The elements of the state space are the system's \textit{states} $s\in\Sigma_\mathcal{S}$, whose exact nature we will leave open for the time being. 

Using the notion of state space, a \textit{property} $\pi$ of a system $\mathcal{S}$ is taken to be an arbitrary subset of $\Sigma_\mathcal{S}$. If the system's state $s$ is an element of a given $\pi$, $s\in\pi\subset\Sigma_\mathcal{S}$, we say that it possesses the corresponding property. A physical system could, for instance, have a liquid and a solid phase. Then, whether the system is liquid is decided based on whether its current state is in the set of all states corresponding to the liquid phase. 

Although this framework applies in more general cases, it may be useful to think of $\Sigma_\mathcal{S}$ as analogous to a system's phase space: a state is then given by the point $x=(p,q)$, where $p$ and $q$ are respectively the generalized momenta and coordinates of the system. As an example, consider the property `having energy less than $E_0$', which corresponds to the set $\pi_{E_0}=\{x|H(x)<E_0\}$, where $H$ is the Hamiltonian function of the system. 

Under set inclusion as ordering relation, the subsets of $\Sigma_\mathcal{S}$ form a lattice, which yields a calculus that can be used to reason about properties of the system. Negation is given by taking the set-theoretic complement $^\perp$, and subset intersection and union correspond, respectively, to the logical `and' ($\wedge$) and `or' ($\vee$). Under complementation, any property $\pi$ is mapped to a property $\pi^\perp$, such that $s$ possesses $\pi^\perp$ whenever it does not possess $\pi$.  For a classical mechanical system, this lattice will be Boolean, ensuring that we can reason about its properties using familiar classical logic. 

By means of this calculus, properties can be combined logically. Thus, if some system e.g. has properties $\pi_1$ and $\pi_2$, it also possesses the property $\pi_{12}=\pi_1\wedge\pi_2$. Likewise, properties can imply one another: a property $\pi_1$ implies $\pi_2$ if it is its subset, and thus, for any state $s$, $s\in\pi_1\rightarrow s\in\pi_2$. Hence, a full specification of all properties of a given system need not be an exhaustive list, but rather, just some minimal set of properties such that all further properties can be derived from it. A system's state is uniquely determined by its properties: any singleton set can be understood as the intersection of sets containing it. Therefore, we can think of the state of a system equivalently as being given by the list of properties the system possesses in that state.  

For each property $\pi$, we stipulate that there exists a measurement $m_\pi$ that determines whether $\mathcal{S}$ possesses $\pi$. That is, measurements are maps $m:\mathcal{S}\to\lbrace0,1\rbrace$ yielding $1$ if $\mathcal{S}$ has property $\pi$, and $0$ if it does not have property $\pi$. The outcome of a measurement for some property $\pi$ is then given by the characteristic function associated to that property. We only consider dichotomic measurements; however, this does not represent a restriction, since any measurement $M$ with multiple outcomes $o_1,o_2,\ldots$ can always be broken down into dichotomic measurements testing for the properties `the value of $M$ is $o_1$', `the value of $M$ is $o_2$', and so on. For each measurement $m$, there exists an \textit{orthogonal measurement} $m^\perp$ corresponding to measuring the property $\pi^\perp$, i.e. the property corresponding to the complementary subset of $\pi$ on the state space $\Sigma_\mathcal{S}$.

Above, the state of a system was defined as a list of all the properties a system may possess in that state. As these properties are in one-to-one correspondence with the possible measurements on a system, this implies that this notion of state is one in which every measurable property possesses a definite value. Such a notion of state is sometimes called \textit{ontic} \cite{Sp2007}, in opposition to an \textit{epistemic} state, which can be viewed as a state of knowledge about the properties of a system---consider the distinction between a phase-space point (ontic) and a Liouville distribution (epistemic). Unless otherwise noted, in the following, a state will always refer to such a complete specification of properties. A quantum state, which does not specify definite values of certain measurable properties, thus does not fall under this notion of state.

A \textit{measurement sequence} $\mu$ is an ordered collection of measurements, $\mu=(m_1,m_2,m_3,\ldots)$. A measurement sequence $\mu$, applied to a state $s$ of $\mathcal{S}$ yields an outcome sequence $\mu(s)$. Since we are only concerned with dichotomic measurements, this sequence will be a string of bits, such that the $n$th bit indicates whether the system possesses property $\pi_n$.

For a classical system, we expect that it is possible to compute the outcomes of every measurement, given sufficient knowledge about the state of the system. This now allows us to state the following \textit{criterion of classicality}: a system $\mathcal{S}$ is \textit{classical}, if it is possible to obtain knowledge of the state $s$ of $\mathcal{S}$ such that the outcomes of all further measurements performed on $\mathcal{S}$ are derivable from this knowledge, that is, if there exists a measurement sequence such that its outcomes suffice to fix the outcomes of all further measurements. Such a measurement sequence may be called a `canonical measurement sequence'.

The purpose of this criterion is simply to capture, in a sufficiently precise way, our intuition that the state of a classical system can be completely known in all of its properties. This is exactly what is denied by the conjunction of postulates \ref{fin} and \ref{add} above: accepting these, a state can never be precisely known, and it will always be possible to perform tests whose outcomes cannot be predicted based on pre-existing knowledge.

To elucidate this notation, we will have another look at the toy theory. The state space $\Sigma_\mathcal{T}$ of a toybit $\mathcal{T}$ is simply the set $\Sigma_\mathcal{T}=\{t_1,t_2,t_3,t_4\}$. Above, we have considered three properties that a toy bit can have, which we may denote $\pi_x$, $\pi_y$ and $\pi_z$. Together with their complements, these properties thus correspond to the following sets: 
\begin{align*}
 &\{t_1,t_2\}= \pi_x;\;\;\{t_3,t_4\}= \pi_x^\perp\\
 &\{t_1,t_3\}= \pi_y;\;\;\{t_2,t_4\}= \pi_y^\perp\\
 &\{t_1,t_4\}= \pi_z;\;\;\{t_2,t_3\}= \pi_z^\perp
\end{align*}

The measurements corresponding to these properties are simply the denoted $m_x$, $m_y$, and $m_z$. Any two of these measurements suffice to infer the outcome of the third, and thus, to fix the entire state. Thus, a possible canonical measurement sequence for the toy bit is
\begin{equation}
\mu=(m_x,m_z^\perp).
\end{equation}
This example also shows that there may be more than one canonical measurement sequence for a given system, which roughly translates to the possibility of different basis expansions of a given state. 

However, by the knowledge balance principle, the state can never be known exactly; in fact, only one of two necessary properties can be known at any given time. 

In the toy theory, then, our condition for classicality would be fulfilled, if we could obtain full knowledge about the ontic state of the system: knowing e.g. that $m_x(t)=0$ and $m_y(t)=1$, we could immediately infer that $m_z(t)=0$, since we know that the state of $\mathcal{T}$ must be $t_3$. If this is not the case, we are subject to an epistemic horizon: it is not possible to know everything that would be necessary to completely describe the system.

With these notions in place, we thus proceed to establish our first main result, showing that not all information about $\mathcal{S}$ can be obtained---more accurately, that there does not exist a set of measurements of $\mathcal{S}$ such that, if their outcomes are known, they suffice to derive all further experimental outcomes on $\mathcal{S}$.

\section{The Uncomputability of Measurement} \label{main}

Our argumentation in the following will make use of the fixed point theorem due to F. W. Lawvere \cite{Law1969}. The original setting of Lawvere's theorem is that of Cartesian closed categories (CCCs), with its most prominent member being the category \textbf{Set}, whose objects are sets, and whose morphisms are maps between them. Yanofsky \cite{Ya2003} provides an elementary introduction to Lawvere's results in the language of set theory. 

Lawvere's result essentially exhibits the `common substructure' behind many of the classical `paradoxes' of self-reference---Cantor's proof of the uncountability of the real numbers \cite{Ca1891}, Russell's paradox that doomed Fregean set theory by exhibiting the construction of a `set of all sets that do not contain themselves' which contains itsef exactly if it does not contain itself \cite{Ru1902}, Tarski's proof of the impossibility of defining a notion of truth in the language of the same system that it pertains to \cite{Ta1936}, Berry's paradox \cite{Ru1908}, the undecidability of the halting problem \cite{Tu1936}, and of course, and perhaps most famously, the incompleteness theorems due to G\"odel \cite{Go1931}.

\subsection{The Existence of Undecidable Measurements}

In the following, we will assume the correspondence proposed in Table~\ref{ana}: that is, physical evolutions correspond to computable processes. In particular, we will study the special case of measurement: acquiring information about another system via, ideally, non-disturbing interactions.

In classical mechanics, such measurements are possible in principle: we are able to simply `read off' information about the properties of a system, without causing any change in these properties. 

Let us thus assume that there exists a computable process associated with every measurement on a given system. That is, for every measurement $m$ there exists an associated program $p$ such that $p$ enumerates the outcomes of the measurement $m$ on every possible state $s$ of $\mathcal{S}$. There thus must exist an onto map between programs and measurements. 

We fix some enumeration for measurements $m_i$, programs $p_j$, and states $s_k$, with $i,j,k\in\mathbb{N}$. Since we are assuming computational dynamics, there can only be denumerably many of each, as there only are denumerably many Turing machines. We will now argue that it is possible to construct a measurement $m_g$ such that there exists an $s_n\in\Sigma_\mathcal{S}$ such that no $p_j$ generates $m_g(s_n)$.

The argument will make use of the method from Ref.~\cite{Ya2003}. We will explicitly construct a measurement $m_g(s_k)$, that is, a function $m_g:\Sigma_\mathcal{S} \to\{0,1\}$, such that it differs from the output for each $p_j$ for at least one $s_k$. 

There exists a function $f(j,k):\mathbb{N}\times\mathbb{N}\to\{0,1\}$ such that it is equal to the output of the $j$th program for the $k$th state. Furthermore, we introduce the arbitrary map $\alpha:\{0,1\}\to\{0,1\}$, and the map $\Delta:\mathbb{N}\to\mathbb{N}\times\mathbb{N}$ that takes $n \in \mathbb{N}$ to the tuple $(n,n)\in\mathbb{N}\times\mathbb{N}$. With these, we construct $g$ as the map that makes the following diagram commute:

\[
\begin{tikzcd}[column sep=large, row sep=huge]
\mathbb{N}\times \mathbb{N} \arrow{r}{f} & \{0,1\} \arrow{d}{\alpha} \\
\mathbb{N} \arrow{u}{\Delta} \arrow{r}{g} & \{0,1\} \\
\end{tikzcd}
\]

The map $g$ constructed in this way then yields sequentially values for a certain measurement, $m_g$, if performed on states of $\mathcal{S}$, i.e. $g(k)=m_g(s_k)$. If $f$ yields the value of every measurement applied to every state, then there must be some $n$ such that $g(k)=f(n,k)$ for all states $s_k$. Choose now $k=n$ and evaluate $g(n)$:

\begin{align}
f(n,n)	& =g(n)\\
		& =\alpha(f(n,n))\nonumber
\end{align}

The first equality is simply our stipulation that $g$ should encode some measurement, and that $f(n,n)$ yields the output of the $n$th program for the $n$th state. The above then shows that the map $\alpha$ must have a fixed point at $f(n,n)$ for the construction to be consistent.

However, we are free in our choice of $\alpha$, and consequently, may choose the negation $\neg(1)=0$, $\neg(0)=1$. But this clearly has no fixed point. But then, this means that no $f$ reproducing every measurement outcome can exist; however, since $f$ yields the outcome of any program applied to any state, there must be some measurement outcome that cannot be produced via a computation. 

The above may be more clear if looked at in terms of Table~\ref{diag1}.

\begin{center}
 \begin{table}[h]
  \caption{Illustration of the fixed-point argument against the possibility of computing every measurement in terms of a diagonalization technique.} 
  
  \begin{tabular}{c | c c c c c c c c}\label{diag1}
                         & $s_1$   & $s_2$    & $s_3$   & $s_4$   & $s_5$   &$\ldots$& $s_n$ &$\ldots$\\ \hline
 $m_1\leftarrow    p_1$  &   (1)   &    0     &    1    &    1    &    1    & $\ldots$&     1    &$\ldots$\\
 $m_2\leftarrow    p_2$  &    1    &   (0)    &    1    &    0    &    0    &         &       0     &\\
 $m_3\leftarrow    p_3$  &    0    &    1     &   (0)   &    0    &    0    &         &       1     &\\
 $m_4\leftarrow    p_4$  &    1    &    0     &    0    &   (1)   &    1    &         &       1     &\\
 $m_5\leftarrow    p_5$  &    0    &    0     &    0    &    1    &   (1)   &         &       0     &\\
                    \vdots             &$\vdots$&          &          &          &         &$\xddots$ & $\vdots$ &\\
 $m_g\leftarrow p_n$ &   0    &     1     &    1    &    0    &    0    &$\ldots$ &    (?)    &$\ldots$\\
                    \vdots             &$\vdots$&          &          &          &         &          & $\vdots$&  $\xddots$\\
  \end{tabular}
 \end{table}
\end{center}

The leftmost column contains the programs $p_i$ and the measurements they implement. The following columns refer to the states $s_k$ of $\mathcal{S}$. The table can be thought of as the table of values of the function $f$: the first row yields $f(1,k)$, i.e. the values produced by the first program for the states of $\mathcal{S}$, the second row yields $f(2,k)$, and so on.

Note that every row also corresponds to a set: the first row corresponds to the set that includes $s_1$, does not include $s_2$, includes $s_3$, etc. Every such assignment of values thus corresponds to a subset of $\Sigma_\mathcal{S}$, and as stipulated, to every such set corresponds a measurement. 

The measurement $m_g$, and its associated set, is then constructed by taking the values on the diagonal (that is, the value the first program assigns to the first state, the second program assigns to the second state, and so on), and inverting them. Thus, we create a measurement that differs from the values produced by the first program in the value assignment to the first state, from the values produced by the second program in the value assigned to the second state, and so on, all down to the diagonal. 

Assuming that there exists some program that produces the values of the measurement thus constructed then yields a contradiction: the value assigned to the state $s_{n}$ is 1, if and only if it is zero. Hence, no program can produce the value of $m_g(s_{n})$.

What, however, happens if we perform the measurement $m_g$, and the system happens to be in state $s_{n}$? If the measurement produces the value $1$, then it follows that it must produce the value $0$; if it produces the value $0$, then it must produce the value $1$. Do we then hover in a state of indecision? Does the measurement simply produce no result?

One might think that this provides a way out; however, it is easy to see that this is not the case. If we allow a measurement to yield `no result', then effectively we have merely added a third outcome. But for three-outcome measurements, a proof exactly analogous to the above one could be constructed to again construct a measurement yielding a contradictory outcome. 

This is analogous to trying to solve the liar-paradox by deciding that `this sentence is false', is neither false nor true, but meaningless. However, then, we can formulate `this sentence is false or meaningless', which must either be true, or not; but if it is true, then it is false or meaningless, yet if it is false or meaningless, then it is true. 

Consequently, we must insist that the measurement does, in fact, produce an outcome. But if this is the case, then, since $s_{n}$ is incompatible with every outcome, upon producing any outcome, the system cannot, in fact, be in state $s_{n}$. Hence, the measurement must produce a change of state.

Furthermore, the system being in the state $s_n$ can neither be said to have the property $\pi_g$ associated to the measurement $m_g$, nor not to have it; yet measurement will always produce a definite judgment. This situation thus bears more than a passing resemblance to the phenomenon of quantum superposition.

This may be compared to trying to build a computer that is able to solve its own halting problem. Recall that the proof of the recursive unsolvability of the halting problem includes the creation of a program that halts if and only if it fails to halt---a contradiction. This is sometimes claimed to make hypercomputation (in any form) impossible; however, this is really only the case for proposed systems capable of solving \textit{their own} halting problem \cite{Or2005}. Thus, an oracle capable of solving the halting problems for Turing machines, but not its own, yields a consistent form of hypercomputation. 

The above, however, resembles a device that is capable of solving its own halting problem: the function $f$ is computable, and thus, it is possible to compute the value of $f(n,n)$, and return the opposite. Hence, this would yield an inconsistency, which can only be avoided if the post-measurement state is different from $s_{n}$. In this sense, the looming inconsistency plays the role of Spekkens' `knowledge balance principle'.

Finally, note that we may view the above equivalently in an `epistemic' way: instead of producing values via measurement, we reinterpret the task as predicting values that will be generated in measurement by computational means. Then, too, we must conclude that there necessarily exist measurements on certain states such that it is impossible to computationally predict their outcome: the criterion of classicality can never be fulfilled. 

In seeming contradiction to the above result, there exist theories that do attribute definite values of properties to quantum systems, nevertheless reproducing the full array of quantum mechanical predictions, such as the de Broglie-Bohm theory \cite{deB1927,bohm1952}. However, the results of Yurtsever \cite{Yu1998}, Bendersky et al. \cite{ben2017}, and Calude and Svozil \cite{CS2008} already imply that any theory capable of reproducing quantum mechanics must be noncomputable.

Nevertheless, to any given particle, Bohmian mechanics associates a full list of properties. In order to see how this is consistent with the above results, it is instructive to further elucidate how the value of each property is arrived at.

Bohmian mechanics, in order to reproduce the predictions of quantum mechanics, must be highly nonlocal. The values of the properties of any single quantum system thus depend on those of every other system. Furthermore, Bohmian mechanics is highly sensitive to initial conditions. In order to reproduce the predictions of quantum mechanics, the initial distribution of particles must satisfy the Born rule;  deviation from this `quantum equilibrium' entails the possibility of nonlocal signaling \cite{val2002}.

Consequently, any randomness observed in the outcomes of measurements in the Bohmian theory must derive from the randomness in the initial conditions. In as much as this is algorithmic randomness, so, too, must the randomness of the initial configuration of particles be. 

But then, it is easy to see why Bohmian mechanics is not a counterexample to, but indeed, serves as an illustration of the above result: due to nonlocality, any given measurement outcome depends, in principle, on the full, algorithmically random, initial conditions (rather than, for instance, merely the configuration contained in the past light cone of a given experiment). In this way, the initial configuration acts as the infinite random $\rho$ in Edis' representation of uncomputable functions; any measurement outcome is thus a function of both this initial random data, and the deterministic algorithm given by the guiding equation.

\subsection{The Existence of Uncomputable Measurement Sequences}

In the preceding subsection, we saw that there is at least one measurement such that, for some given state, its outcome cannot be predicted by computable means (and equivalently, cannot be produced by a computable process).

However, this does not yet yield the sort of phenomena known from quantum mechanics: in particular, it still seems to be possible that there are states such that every measurement performed on them is perfectly well predictable. 

In this section, we will demonstrate another, stronger restriction: that for \textit{every} state $s$, there exist measurement sequences such that not all elements of these sequences can be produced by a computational process.

We will proceed in analogy to the previous section. However, instead of measurements $m_i$, we will now focus on measurement sequences $\mu_i$, $i\in\mathbb{N}$. Again, we will suppose (for contradiction) that every measurement sequence (more accurately, the outcome sequence associated with every measurement sequence for a given, fixed, state $s$) can be generated by some program $p_j$. 

The argument will follow the same lines as the previous one. We explicitly construct an outcome sequence, that is, a function $g:\mathbb{N}\to\{0,1\}$ such that $g(l)=m_l(s)$, where $m_l$ is the $l$th measurement in some measurement sequence $\mu$. Then, we will show that for the constructed $g$, there exists some $n$, such that the value of $g(n)$ cannot be produced by any program $p_j$. 

As before, there is a function $f:\mathbb{N}\times\mathbb{N}\to\{0,1\}$ such that $f(j,k)$ is the value of the $k$th measurement in the measurement sequence $\mu$, produced by the $j$th program (throughout, we leave the state $s$ of $\mathcal{S}$ fixed). 

The map $g$ is then constructed as above. Again, we see that $g(n)$, yielding the value of the $n$th measurement of some measurement sequence $\mu$, must differ from every $f(j,n)$: no program $p_j$ produces the value $g(n)$ as the $n$th measurement of the associated measurement sequence. 

Nevertheless, $g(n)$ corresponds to a valid measurement on $\mathcal{S}$: it is the $n$th measurement of the sequence of measurements obtained by taking the first measurement of the first measurement sequence, the second measurement of the second measurement sequence, and so on, and performing the orthogonal measurement in every case.

Thus, there exists a measurement sequence such that its outcomes cannot be produced by any program $p_j$.

Again, we may lay out the argument in a grid, as in Table~\ref{diag}. 

\begin{center}
 \begin{table}[h]
  \caption{Illustration of the fixed-point argument against the possibility of computing every measurement sequence in terms of a diagonalization technique.} 
  
  \begin{tabular}{c | c c c c c c c c}\label{diag}
  	& $m_1,i$   & $m_2,i$    & $m_3,i$   & $m_4,i$   & $m_5,i$   &$\ldots$& $m_n,i$ &$\ldots$\\ \hline
 $\mu_1(s) \leftarrow    p_1 $  &   (1)   &    0     &   [1]   &    1    &    1    & $\ldots$&     1    &$\ldots$\\
 $\mu_2(s) \leftarrow    p_2 $  &   [1]   &   (0)    &    1    &    0    &    0    &         &       0     &\\
 $\mu_3(s) \leftarrow    p_3 $  &    0    &    1     &   (0)   &    0    &   [0]   &         &       1     &\\
 $\mu_4(s) \leftarrow    p_4 $  &    1    &   [0]    &    0    &   (1)   &    1    &         &       1     &\\
 $\mu_5(s) \leftarrow    p_5 $  &    0    &    0     &    0    &   [1]   &   (1)   &         &       0     &\\
                    \vdots             &$\vdots$&          &          &          &         &$\xddots$ & $\vdots$ &\\
 $\mu_g(s) \leftarrow    p_n $ &   0    &     1     &    1    &    0    &    0    &$\ldots$ &    (?)    &$\ldots$\\
                    \vdots             &$\vdots$&          &          &          &         &          & $\vdots$& $\xddots$\\
 $\mu_{g^\prime}(s) \leftarrow    p_{\beta(n)} $ &   0    &     1     &    0    &    0    &    1    &$\ldots$ &    \ldots    &$\ldots$\\
  \end{tabular}
 \end{table}
\end{center}

Each measurement sequence $\mu_i$ maps to a certain program $p_j$. We construct a new sequence of measurements by taking the measurements yielding the outcomes along the diagonal, and performing the orthogonal measurements. The sequence of outcomes of these measurements must be different from the outcomes obtained from $\mu_1$ in the first element, from those obtained from $\mu_2$ in the second, and so on; hence, no program suffices to derive every measurement outcome produced by that measurement sequence. In particular, trying to produce the measurement outcome of $m_{n}$ produces a contradiction: it yields $0$ if and only if it yields $1$.

One might suppose that this phenomenon is, in some sense, rare. However, this is not the case: we may replace the map $\Delta$ with another, $\langle \beta, \mathrm{Id}\rangle:\mathbb{N}\to\mathbb{N}\times\mathbb{N}$ that takes any $n$ to $(\beta(n),n)$ where $\beta$ is an arbitrary bijection from $\mathcal{N}$ to itself. Each choice of $\beta$ allows us to construct $g^\prime$ such that $\mu_{g^\prime}$ must differ from any computable $\mu$, illustrated in Table~\ref{diag} by using the values in square brackets, and inverting them. But the set of all $\beta$ (the symmetric group over $\mathbb{N}$) has the cardinality of the powerset of $\mathbb{N}$, i.e. $2^{|\mathbb{N}|}$ \cite{kar1956}. Since there are only countably many programs $p_i$, this means that almost all $\mu$ must be uncomputable. Consequently, the computable measurement sequences form a subset of measure zero.

This proof has the form of Richard's paradox \cite{Ri1905}. This paradox involves describing real numbers using natural language. There are some sentences that describe real numbers in the interval $[0,1]$, such as `the number whose decimal expansion is given by the concatenation of all elements of the Fibonacci sequence'. Call these \textit{Richard sentences}, and the numbers they define \textit{Richard numbers}. We can bring them into an ordering, such that we can uniquely refer to `the $n$th digit of the number defined by the $m$th Richard sentence'. Now, consider `the number such that its $n$th digit is equal to 9 minus the $n$th digit of the $n$th Richard number'. Clearly, this is an English sentence describing a real number between $0$ and $1$. However, it cannot be in the previous enumeration: it differs from the first Richard number in the first digit, from the second in the second digit, and so on. 

In our argument, measurement sequences are `named' by programs enumerating their outcomes, and this naming is used to construct a measurement sequence (via its outcome sequence) that cannot be named by---i.e. extracted from---any program, in the same way that we constructed a real number not named by any Richard sentence. Hence, this measurement sequence must differ in at least one measurement from every measurement sequence that can be computed, and thus, cannot be a computable sequence.

Again, we may profitably look at this result in terms of prediction. Then, the programs $p_i$ constitute computations that, given prior knowledge about a system, are used to predict the outcomes of sequences of measurements. The above result then implies that no knowledge exists such that we can computably extract every measurement sequence's outcomes from it. 

Note, however, that this result does not imply that a given measurement's outcome cannot be computably produced by itself, but merely that it cannot be produced \textit{in a given sequence} of measurements. The same measurement could occur in a different position in another measurement sequence, where its outcome may be perfectly computable. 

But this is exactly the behavior we expect in quantum mechanics: for a system in a given state $s$, it might be the case that we can perfectly well predict the outcome of a measurement $m$. If, however, we find that after applying some sequence of measurements $\mu$ to $s$, we can no longer predict $m$, then this must mean that the system is no longer in state $s$---in other words, as above, measurements must be capable of changing the state of a system. A measurement following this state change then may have an entirely different outcome than one preceding it, for instance. In this case, the unpredictable measurement is \textit{complementary} to (at least one of) the measurements that have occurred before.

The crux of the above arguments is the existence of the \textit{diagonal map} map $\Delta:\mathbb{N}\to\mathbb{N}\times\mathbb{N}$. 

Physically, the diagonal map allows the possibility of \textit{cloning}, that is, of creating two identical copies of one and the same system in some state $s$ (recall that a given measurement sequence may contain the complete information of a state; duplicating this information is thus tantamount to duplicating the state itself). It is interesting to note that this map is not realizable in quantum mechanical systems, due to the famed no-cloning theorem \cite{WZ1982}, which can also be thought of as prohibiting exact measurement of every state (for if I could measure every state, I could simply re-prepare it, producing a clone; and if I could clone every state, I could simply measure enough copies to enable tomographic reconstruction of the state).

Interestingly, as noted by Svozil \cite{Sv1995}, there is another way quantum mechanics evades the paradoxical construction. Let us swap the set of classical truth values $\{0,1\}$ for `quantum truth values' $\ket{0}$ and $\ket{1}$. Then, the `negation operator'
\begin{equation}
 D=\ket{0}\!\!\bra{1} + \ket{1}\!\!\bra{0},
\end{equation}
which takes $\ket{0}$ to $\ket{1}$ and vice versa, cannot fulfill the role of the negation in the above proof: the equal superposition 
\begin{equation}
\ket{+}=\frac{1}{\sqrt{2}}(\ket{0}+\ket{1})
\end{equation}
forms a fixed point of the negation operator, yielding $D\ket{+}=\ket{+}$. Consequently, the above proof no longer works---due to the possibility of superposition, the negation operation no longer is fixed-point free.

Taken together, the results in this section imply that, as principle \ref{add} stipulates, additional information is always available about a system---there exist measurements such that their value, for a certain state, cannot be predicted computationally from prior knowledge of that system, and there exist measurement sequences such that they contain measurements that cannot be predicted for \textit{any} state of a system.

\section{Making Undecidability Quantitative} \label{quant}

We have so far established that there are outcomes of measurements on a system $\mathcal{S}$ which cannot be predicted, even given maximal knowledge of the state $s$ of $\mathcal{S}$. Furthermore, we have seen how state change upon measurement is mandated by requirements of consistency, and how measurements may influence the outcomes of measurements following them, leading to a notion of complementarity. 

However, this is not yet fully quantum-like behavior: while we can thus always acquire new information about a system, the characteristically quantum feature is the \textit{loss} or invalidation of old information upon such an information gain---the more accurately one knows a system's position, the less accurate knowledge about its momentum becomes. Moving towards the (epistemic) horizon, discovering new features of the landscape, entails losing sight of areas previously in view. The total amount of information we can gain on a system is bounded, and any new information gained must obey this bound, thus mandating information loss.

The question we will be concerned with in the following is that of \emph{state-space localizability}: how well can we localize $\mathcal{S}$ within its state-space $\Sigma_\mathcal{S}$ by some sequence of measurements? 

Intuitively, it is clear that localizing a system to an ever smaller `area' within its state space requires an increasing amount of information. However, we need to be careful here: in principle, we may define a measurement that asks whether the system is within an arbitrarily small subset of its state space; if this measurement comes out $1$, it seems that we have achieved an arbitrarily fine localization with just a single bit of information.

This, however, is clearly absurd: it is akin to defining a code in which the entire works of Shakespeare are mapped to $1$, and then concluding that they contain only one bit of information. We need, rather, to take into account the information that is needed minimally to describe a given subset, and thus, a property; this then yields a better notion of the information gained in a measurement.

Thus, we appeal to the notion of \textit{Kolmogorov complexity} \cite{Ko1963}. Here, the Kolmogorov or \textit{algorithmic} complexity $K(\xi)$ of a binary string $\xi$ is the length of the shortest program $x$ \cite{Ko1963} producing the string $\xi$ on some universal machine $U$, that is
\begin{equation}
 K(\xi)=\min_{x:U(x)=\xi}|x|,
\end{equation}
where $|x|$ denotes the length of the program $x$. Note that the admissible programs must be \textit{self-delimiting}: no program $x$ can be valid if it contains another valid program as initial substring (that is, there must be something akin to an `end' instruction, terminating bracket, etc.). 

It can be shown that the Kolmogorov complexity is independent of the computing machine $U$ up to a finite factor that roughly quantifies the program simulating $U$ on some different computing machine $U^\prime$; hence, for questions only sensitive to the scaling of this complexity, we can leave the specification of the machine open.

For a given property $\pi$, we can then define its information content as $K(\pi)$---that is, the length of the shortest program that outputs a description of the subset of $\Sigma_\mathcal{S}$ that $\pi$ corresponds to.

Consequently, we can encode the localization of $\mathcal{S}$ into a bit string $\sigma$, such that $n$ bits of $\sigma$ suffice to localize the state within a fraction of $2^{-n}$ of $\Sigma_\mathcal{S}$. If we can obtain information about $\mathcal{S}$ such that we can produce $\sigma$ to an arbitrary degree of precision, then we can localize $\mathcal{S}$ arbitrarily well within its state space. 

We may think about $\sigma$ in the following way: first, divide the state space into two equal subspaces (we will assume a bounded state space for the time being, which is adequate for any real experiment---for instance, a particle is somewhere within the lab, and moving slower than light). If it is found in the right half, the first bit of sigma is equal to $1$, in the left half, it is $0$. Then we proceed to further subdivide the remaining volume, thus yielding a string $\sigma$ such that every bit corresponds to a particular level of a set of nested intervals.

Thus, our question now becomes: Does there exist a measurement sequence $\mu$ such that the associated outcome sequence $\mu(s)$ suffices to compute $\sigma$ to an arbitrary degree of precision? 

In the following, we will answer this question in the negative: there exists a fundamental limit to state-space localizability. To do so, we first note the following: there exists a bit string $\sigma^\prime$ such that finite-length prefixes of $\sigma^\prime$ allow the computation of finite-length prefixes of $\sigma$, and $\sigma^\prime$ is algorithmically random. We may think about $\sigma^\prime$ as a maximally compressed version of $\sigma$.

This follows, because $\sigma$ is equivalent to some outcome sequence $\mu(s)$, and, as shown above, outcome sequences are in general (i.e. with probability $1$) uncomputable. Hence, by the representation theorem of Edis \cite{Edi1998}, there exists an algorithm capable of enumerating bits of $\sigma$ (equivalently, outcomes $\mu(s)$) given a certain random number as an oracle. This random number is $\sigma^\prime$.

If we can now establish that only finitely many bits of $\sigma^\prime$ can be computably generated, then likewise only finitely many bits of $\sigma$ can be generated from it (as otherwise, $\sigma$ would be computable, i.e. there exists a finite program producing infinitely many bits of $\sigma$ from some finite prefix of $\sigma^\prime$). Consequently, any system can only be localized to a finite degree within its state space.

An important property of random sequences is their incompressibility. Intuitively, while a highly regular sequence may admit of a very short description---such as `10,000 times 1'---, the description of a random sequence cannot be significantly shorter than the sequence itself, as it lacks any regularity that could be exploited to this end. 

Then, an infinite string $\xi$ is random if there exists a constant $c$ such that for every $n$-bit prefix of $\xi$, it holds that \cite{Lev1973,Sch1973}
\begin{equation}\label{algrand}
 K(\xi_n)>n-c.
\end{equation}

In accordance with the previous discussion, let $\sigma^\prime$ be an algorithmically random sequence, and let $\sigma^{|m}=C(\sigma^{\prime|n})$, with $n,m<\infty$, for some computation $C$---i.e. let finite prefixes of $\sigma$ be computable from finite prefixes of $\sigma^\prime$. Furthermore, let $\sigma$ encode some outcome sequence $\mu(s)$ such that $\sigma$ enables localization of $s$ within $\Sigma_\mathcal{S}$ to arbitrary precision. Then, only a finite prefix of $\sigma^\prime$ can be produced by some computation, and consequently, only a finite approximation to $\sigma$ can be computable. 

Intuitively, this means that we can localize $\mathcal{S}$ only to a finite degree, no matter what sequence of measurements $\mu$ we perform; or, understood again epistemically, there exists no specification of a system's state that allows arbitrary state-space localizability.

The argument closely follows the method laid out by Chaitin \cite{Ch1992}. 

The key assumption here is again laid out in the correspondence between physical systems, computations, and formal systems (see Table~\ref{ana}): following this analogy, we stipulate that the process by which measurement outcomes are generated is computational. That is, there exists a program such that it enumerates the outcomes of every measurement sequence $\mu$; furthermore, since the bits of $\sigma^\prime$ are likewise computed from measurement outcomes, there must exist some program $p_{\sigma^\prime}$ such that it produces the binary expansion of $\sigma^\prime$. 

Let us then assume, for contradiction, that infinitely many bits of $\sigma^\prime$ can be generated in a computable way. Then, there exists a special-purpose computer $C$ and a program $p$ given by the string
\begin{equation}\label{prog}
 p=\underbrace{00\ldots 01}_{l\mathrm{\,bits}}p_{\sigma^\prime} x,
\end{equation}
such that $C$, executing $p$, does the following: first, it reads the $l$ initial bits, essentially simply telling it the number $l$. Then, $C$ executes the program $p_{\sigma^\prime}$, producing bits of the sequence $\sigma$, not necessarily in any particular order, keeping count of the number $r$ of bits read of $p_{\sigma^\prime}$. It stops this operation as soon as it has generated $r+2l$ bits of $\sigma$. That is, effectively, $C$ needs only execute the program
\begin{equation}\label{prog2}
 p^\prime=\underbrace{00\ldots 01}_{l\mathrm{\,bits}}p_{\sigma^\prime}^{|r} x,
\end{equation}
where $p_{\sigma^\prime}^{|r}$ is that $r$-bit prefix of $p_{\sigma^\prime}$ necessary to generate $r+2l$ bits of $\sigma^\prime$. 

After executing this part of the program, then, $C$ has generated $r+2l$ bits of $\sigma^\prime$. Let $n$ be the index of the rightmost bit it has determined; then, $C$ has generated $r+2l$ bits of the $n$-bit prefix $\sigma^{\prime|n}$ of $\sigma^\prime$. 

If the bits that have been generated are not simply the first $n$ bits of $\sigma^\prime$, then there are $n-r-2l$ bits in $\sigma^{\prime|n}$ that have not yet been generated by $C$. These bits are given to $C$ as the string $x$ in the above program. Consequently, the action of $C$ on $p^\prime$ is 
\begin{equation}
 C(p^\prime)=\sigma^{\prime|n};
\end{equation}
that is, the program $p^\prime$ enables $C$ to compute an $n$-bit prefix of $\sigma^\prime$. 

Now, the total length of the program $p^\prime$ is given by
\begin{equation}
 |p^\prime|=l+r+n-r-2l=n-l.
\end{equation}

This allows us to put an upper bound on the Kolmogorov complexity of $\sigma^{|n}$ of
\begin{equation}
 K(\sigma^{\prime|n})\leq n-l+c^\prime,
\end{equation}
where $c^\prime$ is the length of a program simulating $C$ on some universal machine $U$, since we know that there exists a program of length $n-l$ for $C$ computing $\sigma^{\prime|n}$. 

The randomness of $\sigma^\prime$ now implies that (see Eq.~\ref{algrand})
\begin{equation}
 n-c < K(\sigma^{\prime|n}) \leq n-l+c^\prime,
\end{equation}
which can only hold if $l<c+c^\prime$. But $l$ is a free parameter in $p^\prime$; consequently, we can choose $l=c+c^\prime$, and reach a contradiction: if we can obtain $r+2l$ bits of $\sigma^\prime$, then $K(\sigma^{\prime|n})\leq n-c$; but this is impossible due to the randomness of $\sigma^{\prime|n}$. Thus, we can only obtain less than $r+2l$ bits of $\sigma^{\prime|n}$. 

Consequently, once we have performed enough measurements such that their outcomes suffice to generate $r+2l$ bits of $\sigma^\prime$, further measurements cannot yield additional information increasing the localizability of $\mathcal{S}$. The reason for this is that, by the results of Section~\ref{main}, the sequence of measurement outcomes localizing $s$ in $\Sigma_\mathcal{S}$ must be uncomputable. But every uncomputable sequence can be obtained via a computation that has a random string as an oracle. However, only finite prefixes of algorithmically random sequences can be generated computationally. Hence, putting this together, any system can only be located within its state-space up to finite accuracy. 

Now, if this bound is attained, there are certainly measurements that ought to be able to increase our knowledge about $\mathcal{S}$, and consequently, help localize it further within $\Sigma_\mathcal{S}$. The result in the previous section guarantees as much: there must be a measurement yielding information not reducible to that obtained via the sequence of measurements performed so far.

This situation is comparable to what we find in Spekkens' toy theory: having obtained a certain amount of information, another measurement is in principle possible which would further increase our knowledge. However, there exists a bound on this knowledge---given, in the case of the toy theory, by the knowledge balance principle. 

One possibility now would be that the measurement necessarily fails---that whenever we are in a position to increase our knowledge beyond the established bounds, suddenly our equipment malfunctions, or the measurement otherwise does not yield a result. However, this seems absurd: after all, these experiments are perfectly feasible, as long as our knowledge remains within the limits. Furthermore, as we have seen, measurements that fail to produce an outcome do not help to avert the consequences of these arguments.

The other possibility is that the measurements succeed, and still, the knowledge bounds are not exceeded. This necessitates, however, that prior information must become obsolete: that is, that knowledge gathered by previous measurements no longer applies to the system. 

This is now exactly what we find in quantum mechanics: in order not to violate the uncertainty principle, gaining accuracy in our knowledge of, say, a system's position entails an equivalent loss of knowledge about its momentum. The volume within which we can maximally localize a system remains constant; we can at best deform it. We have already seen that it must be in principle possible for a measurement to change the state of the system; the result in this section then imposes a constraint on how this change occurs---namely, in accordance with the bound on knowledge that can be attained about a system.

At this point, one might well ask why, if we cannot localize the system further via computable means, additional measurement outcomes are not then produced via noncomputable processes. After all, we have already seen that the dynamics cannot be fully computational. 

The answer to this question was already given in the previous section: as we saw, it is not consistently possible to yield such `undecidable' outcomes and remain in the same state. Consequently, a state change must occur once we reach the epistemic horizon, such that the measurement outcome we obtain could be derived from the post-measurement state via computational means. Thus, the bound is upheld, and we obtain no further localizing information.

The noncomputational processes thus seem to, in some sense, cover their tracks, so as to not lead to inconsistency. One might compare this to how, in certain interpretations of quantum mechanics (such as the de Broglie-Bohm one), faster-than-light influences are barred from producing causal paradoxes, enabling the `peaceful coexistence' \cite{shi1978} of quantum mechanics and special relativity.

This amounts to a proof of the first of the two principles above: by computational means, we can only obtain a finite amount of information about the system $\mathcal{S}$, sufficient to localize it in its state space to within a finite smallest volume.

Note that this is not necessarily a restriction on the knowability of a system's state. One might hold---against the definition of state given above---that a valid and complete state of a system is not obliged to yield definite values for every conceivable measurement. Indeed, according to views of the quantum state that grant it ontic status, this is exactly the case in quantum mechanics. In such a case, the state of the system can be perfectly known---however, this state itself does not perfectly localize the system within its state space. 

\section{Conclusion} \label{conc}

Special relativity can be obtained from constraints on the capacities of observers to perform certain measurements. Already Bohr noted a kinship to quantum mechanics in this regard \cite{bohr1939}:

\begin{quotation}
The impossibility of an unambiguous separation between space and time without reference to the observer, and the impossibility of a sharp separation between the behaviour of objects and their interaction with the means of observation are, in fact, straightforward consequences of the existence of a maximum velocity of propagation of all actions and of a minimum quantity of any action, respectively. 
\end{quotation}

Einstein's thought experiments placed the observer's instruments within the purview of the physics they sought to describe, forcing them to work with real clocks and rulers, subject to limitations such as the finiteness of the speed of light. 

Our arguments can be read in the same operational spirit. Limitations exist for measurement apparatuses precisely due to their being embedded within the same physical context as the systems they observe. Just as the constraints of special relativity imply the impossibility of a universal standard of simultaneity and the nonexistence of an absolute length scale, leading to the observable phenomena of time dilation and length contraction, our results imply restrictions on the amount of knowledge available about an object system, leading to many of the phenomena of quantum mechanics (as detailed in the various reconstructions \cite{Rov1996,BZ2003,Fu2002,MM2013,HW2015,vW1985,Sp2007}).

We thus started our investigation by noting the role of two postulates, a postulate of finite information and a postulate of additional information, in derivations of the formalism of quantum mechanics. We identified self-reference, or, more accurately, the undecidability of certain measurement outcomes due to self-reference, as a common thread behind these postulates. Thus, the second postulate was derived as the impossibility of predicting all possible measurement outcome sequences from the knowledge of a system's state in Section~\ref{main}, and the first postulate as the impossibility of localizing a system in its state space to arbitrary accuracy in Section~\ref{quant}. As a consequence, the classical ideal of perfect knowledge and predictivity is as unachievable as Hilbert's dream of finding a single, finite axiomatization of all of mathematics.

This provides an example of what Svozil \cite{Sv1996} has termed the `third path' in dealing with the limits of knowledge as imposed by physical or logical constraints: rather than excluding them as mere `artifacts' or, at the opposite end, assuming a stance of defeatism towards such unknowables, one ought to instead investigate their implications towards theory building in a systematic manner.

Hence, in any theory meeting the requirements of these proofs---that is, any theory which allows for the instantiation of the maps used in the proofs by means of some physical process---, these `postulates' need not be postulated at all; rather, they come `for free', and with them, all the quantum effects that they entail. In particular, this includes classical mechanics: since it allows for universal computation, and for the in-principle arbitrary determination of system states, all of the maps used can be instantiated. But then, quantum mechanics (or at least, those elements of it entailed by the epistemic horizons) is not separate from classical mechanics after all; rather, it is merely the consistent completion of the former in regimes where the theory would otherwise lead to paradoxical self-reference. 

In the same sense, special relativity is the consistent completion of Newtonian mechanics in regimes of velocities comparable to that of light: just as we find novel observable phenomena when approaching this universal speed limit, novel phenomena manifest themselves when we attempt to achieve perfect knowledge of the state of an object system. In both cases, the classical formalism becomes deformed---in the mathematical sense---by the presence of an invariant parameter. Just as we do not ordinarily move at velocities comparable to that of light, we do not ordinarily have even close to full knowledge of the state of any given object---think of the myriad ways in which the atoms of your desk could be arranged, without you noticing the slightest difference. Thus, it is only in the regimes where we are in a position to obtain close to maximal knowledge of a given system's state---which typically correspond to very tightly controlled systems of few microscopic constituents---that the deformation's effects become observable. 

This opens up a new avenue of investigation into Wheeler's perennial question for the origins of quantum mechanics. As we have seen, Wheeler himself at times pursued the same path. Following it thus seems to hold great potential for answering otherwise difficult questions about quantum mechanics. 

We have shown that, in general, there are always measurement outcomes that cannot be predicted; moreover, there are properties such that a system in a certain state can neither be said to possess nor not to possess them, mirroring the phenomenon of superposition. 

Furthermore, whether these outcomes can be predicted depends on the sequence of measurements that have already been performed. Thus, a given measurement might be perfectly predictable in one sequence---within one measurement context---yet become unpredictable in another. This yields the central quantum phenomenon of complementarity: certain measurements may influence the value of subsequently performed ones. Moreover, we see how measurement necessarily affects the state of a system.

Finally, the impossibility of localizing a system to an arbitrary degree within its state space yields the uncertainty principle: there exists a fixed maximum knowledge that can be extracted about a system's state. 

We thus approach quantum phenomena from two different angles: one, we have seen how the limitations imposed upon measurement by the arguments given above can serve as a justification of the epistemic horizons employed as foundational principles for quantum mechanics in various reconstructions; and two, we gave explicit examples of how phenomena closely paralleling the quantum phenomena of superposition, complementarity, and the uncertainty principle emerge directly from these restrictions.

Nevertheless, I do not claim to have achieved a reconstruction of the full quantum formalism. While the above results appear very suggestive, it remains to be seen whether the entirety of quantum phenomena can be derived from these considerations. 

Should this be the case, however, there emerges a picture of quantum evolution that is overall uncomputable, but can, as in Edis' representation theorem, be decomposed into a deterministic and a random part---the randomness taking over whenever knowledge about the system's state reaches the epistemic horizon. 

Such a `hypercomputational interpretation' has intriguing consequences for the measurement problem. Usually, this is stated as the discrepancy between von Neumann's `process I' and `process II', the discontinuous `collapse' of the wave function to an eigenstate of the measurement operator and the deterministic, unitary Schrödinger dynamics, respectively \cite{VN1955}. The present picture naturally leads to such a two-tiered dynamics: whenever we would otherwise exceed the epistemic boundaries, randomness takes over.

But this two-tieredness is ultimately merely apparent: in the end, there exists only a single physical evolution, which, however, cannot be computable. Nevertheless, any observer having a certain amount of information about a system---and thus, being appropriately correlated with it---will be forced to model this dynamics as a certain random event. In this sense, the present view is inherently \textit{relational}: whether smooth Schrödinger or discontinuous measurement dynamics have to be applied depends on the information one system has about another.

Thus, the study of the role of undecidability and self-reference in the foundations of quantum mechanics seems a promising route for future research. One interesting possibility is that, in certain cases, undecidability appears to be a resource that can be used to outperform classical capacities: Van den Nest and Briegel have shown that, in the setting of measurement-based quantum computation, a quantum state yields a speedup whenever the logic derived from the graph representing the state is undecidable \cite{VB2008}.

The method of proof used in arriving at our results derives from a fixed-point theorem due to Lawvere \cite{Law1969}. 

The specific setting of Lawvere's theorem is that of Cartesian closed categories, whose most prominent representative is the category \textbf{Set}. The critical feature of these categories that enables the proof of this theorem is the existence of a diagonal map $\Delta$. Basically, this map enables the copying of information, in our case of the outcome sequence obtained by applying a sequence of measurements to a state (which, in a classical world, is equivalent to the state itself). 

This is, of course, an operation that is expressly forbidden in quantum mechanics---the cloning of information, or copying of quantum states \cite{WZ1982}, respectively the possibility of perfect state discrimination. As Baez \cite{Ba2004} notes, one of the features that distinguishes the category \textbf{Hilb}, whose objects are Hilbert spaces and whose morphisms are given by linear operators, from \textbf{Set} is in fact the absence of such a cloning operation, due to the fact that it is not Cartesian closed. Indeed, as he and Stay suggest \cite{BS2011}, `Cartesian' can be taken to be the antonym of `quantum' in the category theoretical setting. 

A possible implication of this is then that quantum theory appears strange and in need of `interpretation' to us only because of this `categorical mismatch', trying to apply the logic of sets to objects that are not well described in this way. At first, it seems to remain mysterious why we would do so, however. Baez \cite{Ba2004} suggests that this may be the case because the objects of our everyday experience seem to be well approximated as `elements of sets'. But this is itself a fact in need of explanation. 

A different suggestion might be that this is precisely due to the fact that Cartesian closed categories, unlike $\mathbf{Hilb}$, allow for the copying of information. That this is a feature essential to our apprehension of the world becomes clear if we consider that without it, communication in the everyday sense is impossible: such communication has only taken place if afterwards all involved parties possess the same information. Thus, in one way or another, this feature is present in everything that we can talk about, write about, or even reason about; hence, it is impressed on the way we grasp the world, and underlies every model we construct of it. 

\section*{Acknowledgements}

My first and foremost thanks is due to Dagmar Bruß and Hermann Kampermann, whose guidance and tutelage I had the great privilege to receive, and who have been instrumental in the sharpening of the ideas presented here. Furthermore, I wish to thank Karl Svozil and Noson Yanofsky for invaluable discussion of the material compiled in this article.

\bibliography{EpiHor.bib}

\begin{thebibliography}{10}

\bibitem{bar2004}
John~D Barrow, Paul~CW Davies, and Charles~L Harper~Jr.
\newblock {\em {Science and ultimate reality: Quantum theory, cosmology, and
  complexity}}.
\newblock Cambridge University Press, Cambridge, 2004.

\bibitem{Gr2003}
Alexei Grinbaum.
\newblock {Elements of information-theoretic derivation of the formalism of
  quantum theory}.
\newblock {\em International Journal of Quantum Information}, 1(03):289--300,
  2003.

\bibitem{Rov1996}
Carlo Rovelli.
\newblock {Relational quantum mechanics}.
\newblock {\em International Journal of Theoretical Physics}, 35(8):1637--1678,
  1996.

\bibitem{BZ2003}
{\v{C}}aslav Brukner and Anton Zeilinger.
\newblock {Information and fundamental elements of the structure of quantum
  theory}.
\newblock In {\em Time, quantum and information}, pages 323--354. Springer,
  2003.

\bibitem{Fu2002}
Christopher~A Fuchs.
\newblock {Quantum mechanics as quantum information (and only a little more)}.
\newblock {\em arXiv preprint quant-ph/0205039}, 2002.

\bibitem{MM2013}
Llu{\'\i}s Masanes, Markus~P M{\"u}ller, Remigiusz Augusiak, and David
  P{\'e}rez-Garc{\'\i}a.
\newblock {Existence of an information unit as a postulate of quantum theory}.
\newblock {\em Proceedings of the National Academy of Sciences},
  110(41):16373--16377, 2013.

\bibitem{HW2015}
Philipp~Andres H{\"o}hn and Christopher~SP Wever.
\newblock {Quantum theory from questions}.
\newblock {\em Physical Review A}, 95(1):012102, 2017.

\bibitem{vW1985}
Carl~Friedrich {von Weizs{\"a}cker}, Thomas G{\"o}rnitz, and Holger Lyre.
\newblock {\em {The structure of physics}}.
\newblock Springer, 2006.

\bibitem{Sp2007}
Robert~W Spekkens.
\newblock {Evidence for the epistemic view of quantum states: A toy theory}.
\newblock {\em Physical Review A}, 75(3):032110, 2007.

\bibitem{CZ2012}
Thomas~L Curtright and Cosmas~K Zachos.
\newblock {Quantum mechanics in phase space}.
\newblock {\em Asia Pacific Physics Newsletter}, 1(01):37--46, 2012.

\bibitem{Gr2005}
Alexei Grinbaum.
\newblock {Information-Theoretic Princple Entails Orthomodularity of a
  Lattice}.
\newblock {\em Foundations of Physics Letters}, 18(6):563--572, 2005.

\bibitem{Bir1936}
Garrett Birkhoff and John {von Neumann}.
\newblock {The logic of quantum mechanics}.
\newblock {\em Annals of mathematics}, pages 823--843, 1936.

\bibitem{Ch1990}
Gregory~J. Chaitin.
\newblock {Undecidability and randomness in pure mathematics}.
\newblock In {\em Information, Randomness \& Incompleteness}, pages 307--313.
  World Scientific, Singapore, 1990.

\bibitem{Yu1998}
Ulvi Yurtsever.
\newblock {Quantum mechanics and algorithmic randomness}.
\newblock {\em Complexity}, 6(1):27--34, 2000.

\bibitem{ben2017}
Ariel Bendersky, Gabriel Senno, Gonzalo de~la Torre, Santiago Figueira, and
  Antonio Acin.
\newblock {Nonsignaling Deterministic Models for Nonlocal Correlations have to
  be Uncomputable}.
\newblock {\em Physical review letters}, 118(13):130401, 2017.

\bibitem{CS2008}
Cristian~S Calude and Karl Svozil.
\newblock {Quantum randomness and value indefiniteness}.
\newblock {\em Advanced Science Letters}, 1(2):165--168, 2008.

\bibitem{KS1967}
Simon Kochen and Ernst~P Specker.
\newblock {The problem of hidden variables in quantum mechanics}.
\newblock {\em Journal of Mathematics and Mechanics}, (17):59--87, 1969.

\bibitem{Fe1982}
Richard~P Feynman.
\newblock {Simulating physics with computers}.
\newblock {\em International journal of theoretical physics}, 21(6-7):467--488,
  1982.

\bibitem{Edi1998}
Taner Edis.
\newblock {How G{\"o}del's Theorem Supports the Possibility of Machine
  Intelligence}.
\newblock {\em Minds and Machines}, 8(2):251--262, 1998.

\bibitem{Ch1975}
Gregory~J Chaitin.
\newblock {A theory of program size formally identical to information theory}.
\newblock {\em Journal of the ACM (JACM)}, 22(3):329--340, 1975.

\bibitem{Tu1936}
Alan~Mathison Turing.
\newblock {On computable numbers, with an application to the
  Entscheidungsproblem}.
\newblock {\em Journal of Mathematics}, 58(345-363):5, 1936.

\bibitem{Ca2001}
Cristian~S Calude, Peter~H Hertling, Bakhadyr Khoussainov, and Yongge Wang.
\newblock {Recursively enumerable reals and Chaitin $\Omega$ numbers}.
\newblock In {\em Annual Symposium on Theoretical Aspects of Computer Science},
  pages 596--606. Springer, 1998.

\bibitem{Sv1993}
Karl Svozil.
\newblock {\em {Randomness and Undecidability in Physics}}.
\newblock World Scientific, Singapore, 1993.

\bibitem{svo2018}
Karl Svozil.
\newblock {\em {Physical (A)Causality}}, volume 192 of {\em Fundamental
  Theories of Physics}.
\newblock Springer International Publishing, Cham, Heidelberg, New York,
  Dordrecht, London, 2018.

\bibitem{Go1931}
Kurt G{\"o}del.
\newblock {{\"U}ber formal unentscheidbare S{\"a}tze der Principia Mathematica
  und verwandter Systeme I}.
\newblock {\em Monatshefte f{\"u}r Mathematik und Physik}, 38(1):173--198,
  1931.

\bibitem{Ho1980}
William~A Howard.
\newblock {The formulae-as-types notion of construction}.
\newblock {\em To HB Curry: essays on combinatory logic, lambda calculus and
  formalism}, 44:479--490, 1980.

\bibitem{Be1973}
Charles~H Bennett.
\newblock {Logical reversibility of computation}.
\newblock {\em IBM journal of Research and Development}, 17(6):525--532, 1973.

\bibitem{Ch1982}
Gregory~J Chaitin.
\newblock {G{\"o}del's theorem and information}.
\newblock {\em International Journal of Theoretical Physics}, 21(12):941--954,
  1982.

\bibitem{Cu2015}
Toby~S Cubitt, David Perez-Garcia, and Michael~M Wolf.
\newblock {Undecidability of the spectral gap}.
\newblock {\em Nature}, 528(7581):207--211, 2015.

\bibitem{Be1966}
Robert Berger.
\newblock {\em {The undecidability of the domino problem}}.
\newblock Number~66. American Mathematical Society, 1966.

\bibitem{Ll1993}
Seth Lloyd.
\newblock {Quantum-mechanical computers and uncomputability}.
\newblock {\em Physical Review Letters}, 71(6):943, 1993.

\bibitem{Ll1994}
Seth Lloyd.
\newblock Necessary and sufficient conditions for quantum computation.
\newblock {\em Journal of Modern Optics}, 41(12):2503--2520, 1994.

\bibitem{Ei2012}
Jens Eisert, Markus~P M{\"u}ller, and Christian Gogolin.
\newblock {Quantum measurement occurrence is undecidable}.
\newblock {\em Physical review letters}, 108(26):260501, 2012.

\bibitem{Pop1950}
Karl~R Popper.
\newblock {Indeterminism in quantum physics and in classical physics. Part I}.
\newblock {\em The British Journal for the Philosophy of Science},
  1(2):117--133, 1950.

\bibitem{Rot1964}
Jerome Rothstein.
\newblock {Thermodynamics and some undecidable physical questions}.
\newblock {\em Philosophy of Science}, 31(1):40--48, 1964.

\bibitem{Fu2016}
Christopher~A Fuchs.
\newblock {On participatory realism}.
\newblock In {\em Information and Interaction}, pages 113--134. Springer, 2017.

\bibitem{whe1974}
John Wheeler.
\newblock {Add "Participant" to "Undecidable Propositions" to arrive at
  Physics} [online].
\newblock 1974.
\newblock URL:
  \url{https://jawarchive.files.wordpress.com/2012/03/twa-1974.pdf} [cited
  2018-09-16].

\bibitem{ber1991}
J.~Bernstein.
\newblock {\em {Quantum Profiles}}.
\newblock Princeton University Press, Princeton, NJ, 1991.

\bibitem{DC1977}
Maria~Luisa Dalla~Chiara.
\newblock {Logical self reference, set theoretical paradoxes and the
  measurement problem in quantum mechanics}.
\newblock {\em Journal of Philosophical Logic}, 6(1):331--347, 1977.

\bibitem{Br1995}
Thomas Breuer.
\newblock {The impossibility of accurate state self-mea\-sure\-ments}.
\newblock {\em Philosophy of Science}, pages 197--214, 1995.

\bibitem{Bre1999}
Thomas Breuer.
\newblock {J}ohn von {N}eumann met {K}urt {G}{\"o}del: Undecidable statements
  in quantum mechanics.
\newblock In Maria Luisa~Dalla Chiara, Roberto Giuntini, and Federico Laudisa,
  editors, {\em {Language, Quantum, Music: Selected Contributed Papers of the
  Tenth International Congress of Logic, Methodology and Philosophy of Science,
  {F}lorence, August 1995}}, pages 159--170. Springer Netherlands, Dordrecht,
  1999.

\bibitem{Ae2005}
Sven Aerts.
\newblock {Undecidable classical properties of observers}.
\newblock {\em International Journal of Theoretical Physics},
  44(12):2113--2125, 2005.

\bibitem{MZ1978}
Martin Zwick.
\newblock {Quantum measurement and G\"odel's proof}.
\newblock {\em Speculations in Science and Technology}, 1(2):I978, 1978.

\bibitem{Per1982}
Asher Peres and Wojciech~H Zurek.
\newblock {Is quantum theory universally valid?}
\newblock {\em American Journal of Physics}, 50(9):807--810, 1982.

\bibitem{Bru2010}
{\v{C}}aslav Brukner.
\newblock {Quantum complementarity and logical indeterminacy}.
\newblock {\em Natural Computing}, 8(3):449--453, 2009.

\bibitem{Pat2010}
Tomasz Paterek, Johannes Kofler, Robert Prevedel, Peter Klimek, Markus
  Aspelmeyer, Anton Zeilinger, and {\v{C}}aslav Brukner.
\newblock {Logical independence and quantum randomness}.
\newblock {\em New Journal of Physics}, 12(1):013019, 2010.

\bibitem{CJ2005}
Cristian~S Calude and Helmut J{\"u}rgensen.
\newblock {Is complexity a source of incompleteness?}
\newblock {\em Advances in Applied Mathematics}, 1(35):1--15, 2005.

\bibitem{CS2007}
Cristian~S Calude and Michael~A Stay.
\newblock {From Heisenberg to G{\"o}del via Chaitin}.
\newblock {\em International Journal of Theoretical Physics}, 46(8):2013--2025,
  2007.

\bibitem{Law1969}
F~William Lawvere.
\newblock {Diagonal arguments and cartesian closed categories}.
\newblock In {\em Category theory, homology theory and their applications II},
  pages 134--145. Sprin\-ger, 1969.

\bibitem{Ya2003}
Noson~S Yanofsky.
\newblock {A universal approach to self-referential paradoxes, incompleteness
  and fixed points}.
\newblock {\em Bulletin of Symbolic Logic}, 9(03):362--386, 2003.

\bibitem{Ca1891}
Georg Cantor.
\newblock {{\"U}ber eine elementare Frage der Mannigfaltigkeitslehre}.
\newblock {\em Jahresbericht der Deutschen Mathematiker-Vereinigung}, 1:75--78,
  1892.

\bibitem{Ru1902}
Bertrand Russell.
\newblock {Letter to Frege}.
\newblock In Jean van Heijenoort, editor, {\em From Frege to G{\"o}del: a
  source book in mathematical logic, 1879-1931}. Harvard University Press,
  Cambridge, 1967.

\bibitem{Ta1936}
Alfred Tarski.
\newblock {Der Wahrheitsbegriff in den formalisierten Sprachen}.
\newblock {\em Studia Philosophica}, 1:261--405, 1936.

\bibitem{Ru1908}
Bertrand Russell.
\newblock {Mathematical logic as based on the theory of types}.
\newblock {\em American journal of mathematics}, 30(3):222--262, 1908.

\bibitem{Or2005}
Toby Ord and Tien~D Kieu.
\newblock {The diagonal method and hypercomputation}.
\newblock {\em The British journal for the philosophy of science},
  56(1):147--156, 2005.

\bibitem{deB1927}
Louis De~Broglie.
\newblock La m{\'e}canique ondulatoire et la structure atomique de la
  mati{\`e}re et du rayonnement.
\newblock {\em J. Phys. Radium}, 8(5):225--241, 1927.

\bibitem{bohm1952}
David Bohm.
\newblock A suggested interpretation of the quantum theory in terms of" hidden"
  variables, i and ii.
\newblock {\em Physical review}, 85(2):166, 1952.

\bibitem{val2002}
Antony Valentini.
\newblock Signal-locality in hidden-variables theories.
\newblock {\em Physics Letters A}, 297(5-6):273--278, 2002.

\bibitem{kar1956}
Abraham Karrass and D~Solitar.
\newblock {Some remarks on the infinite symmetric groups}.
\newblock {\em Mathematische Zeitschrift}, 66(1):64--69, 1956.

\bibitem{Ri1905}
Jules Richard.
\newblock {Les Principes des Mathématiques et le Problème des Ensembles}.
\newblock In Jean van Heijenoort, editor, {\em From Frege to G{\"o}del: a
  source book in mathematical logic, 1879-1931}. Harvard University Press,
  Cambridge, 1967.

\bibitem{WZ1982}
William~K Wootters and Wojciech~H Zurek.
\newblock {A single quantum cannot be cloned}.
\newblock {\em Nature}, 299(5886):802--803, 1982.

\bibitem{Sv1995}
Karl Svozil.
\newblock {A constructivist manifesto for the physical sciences—Constructive
  re-interpretation of physical undecidability}.
\newblock In {\em The Foundational Debate}, pages 65--88. Springer, Dordrecht,
  1995.

\bibitem{Ko1963}
Andrei~N Kolmogorov.
\newblock {On tables of random numbers}.
\newblock {\em Sankhy{\=a}: The Indian Journal of Statistics, Series A}, pages
  369--376, 1963.

\bibitem{Lev1973}
Leonid Levin.
\newblock {On the notion of a random sequence}.
\newblock {\em Soviet Mathematics Doklady}, 14:1413--1416, 1973.

\bibitem{Sch1973}
Claus-Peter Schnorr.
\newblock {Process complexity and effective random tests}.
\newblock {\em Journal of Computer and System Sciences}, 7(4):376--388, 1973.

\bibitem{Ch1992}
Gregory~J Chaitin.
\newblock {Information-theoretic incompleteness}.
\newblock {\em Applied Mathematics and Computation}, 52(1):83--101, 1992.

\bibitem{shi1978}
Abner Shimony.
\newblock {Metaphysical problems in the foundations of quantum mechanics}.
\newblock {\em International Philosophical Quarterly}, 18(1):3--17, 1978.

\bibitem{bohr1939}
Niels Bohr.
\newblock The causality problem in atomic physics.
\newblock {\em New theories in physics}, pages 11--30, 1939.

\bibitem{Sv1996}
Karl Svozil.
\newblock {Undecidability Everywhere?}
\newblock In John~L Casti and Anders Karlqvist, editors, {\em Boundaries And
  Barriers: On The Limits To Scientific Knowledge}, pages 215--237. Basic
  Books, New York, 1996.

\bibitem{VN1955}
John von Neumann.
\newblock {\em {Mathematical Foundations of Quantum Mechanics}}.
\newblock Number~2. Princeton University Press, 1955.

\bibitem{VB2008}
Maarten Van~den Nest and Hans~J Briegel.
\newblock {Measurement-based quantum computation and undecidable logic}.
\newblock {\em Foundations of Physics}, 38(5):448--457, 2008.

\bibitem{Ba2004}
John~C Baez.
\newblock {Quantum quandaries: a category-theoretic perspective}.
\newblock In Dean~P Rickles, Stephen~R French, and Juha~T Saatsi, editors, {\em
  Structural Foundations of Quantum Gravity}, pages 240--267. Oxford University
  Press, Oxford, 2006.

\bibitem{BS2011}
John Baez and Mike Stay.
\newblock {Physics, topology, logic and computation: a Rosetta Stone}.
\newblock In {\em New structures for physics}, pages 95--172. Springer, Berlin,
  2010.

\end{thebibliography}

\end{document}